\documentclass[12pt,preprint]{aastex}

\usepackage[usenames]{color}

\slugcomment{Not to appear in Nonlearned J., 45.}

\shorttitle{Evolutionary transition between compact stars}
\shortauthors{Nishimura et al.}

\begin{document}

\title{$R$-PROCESS NUCLEOSYNTHESIS IN MHD JET EXPLOSIONS OF CORE-COLLAPSE SUPERNOVAE}

\author{Sunao Nishimura\altaffilmark{1}, Kei Kotake\altaffilmark{2}, 
Masa-aki Hashimoto\altaffilmark{1}, Shoichi Yamada\altaffilmark{3}}
%\affil{Department of Physics, Kyushu University, Fukuoka 810-8560, Japan}
%\email{hashi@gemini.rc.kyushu-u.ac.jp}

\and

\author{Nobuya Nishimura\altaffilmark{1}, Shinichiro Fujimoto\altaffilmark{4},
Katsuhiko Sato\altaffilmark{2,5}}
%\affil{Department of Physics, Hokkaido University, Sapporo, 060-8810, Japan}
%\email{fujimoto@phys.hokudai.ac.jp}

\altaffiltext{1}{Department of Physics, Kyushu University, Ropponmatsu, Fukuoka 810-8560, Japan.}
\altaffiltext{2}{Department of Physics, School of Science, University of Tokyo, 7-3-1
Hongo, Tokyo 113-0033, Japan.}
\altaffiltext{3}{Science and Engineering, Waseda University, 3-4-1 Okubo, Tokyo 169-8555, Japan.}
\altaffiltext{4}{Kumamoto National College of Technology, Kumamoto 861-1102, Japan.}
\altaffiltext{5}{Research Center for the Early Universe, University of Tokyo, 7-3-1
Hongo, Tokyo 113-0033, Japan.}

\begin{abstract}
We investigate $r$-process nucleosynthesis during the magnetohydrodynamical (MHD) explosion of supernova in a massive star of 13 $M_{\odot}$. 
Contrary to the case of the spherical explosion, jet-like explosion due
to the combined effects of the rotation and magnetic field
 lowers the electron fraction significantly inside the layers above the iron core. We find that the
ejected material of low electron fraction responsible for the $r$-process comes out from the silicon
rich layer of the presupernova model.
This leads to the production up to the third peak in the solar $r$-process elements.
We examine whether the fission affects the $r$-process paths by 
using the full nuclear reaction network with both the spontaneous and $\beta$-delayed fission included.
Moreover, we pay particular attention how the mass formula affects the $r$-process peaks
with use of two mass formulae. 
It is found that both formulae can reproduce the global abundance pattern up to the third peak
though detailed distributions are rather different.
We point out that there are variations in the $r$-process nucleosynthesis
if the MHD effects play an important role in the supernova explosion.
\end{abstract}


\section{INTRODUCTION}

It has been considered that the origin of heavy neutron-rich elements like uranium is mainly due to the $r$-process
nucleosynthesis that occurs during the supernova explosions and/or neutron star
mergers \citep{th01,qi03}. 
The main issue concerning the $r$-process research is to reproduce
the three peaks ($A\simeq$ 80, 130, and 195) in the abundance pattern for the $r$-elements
in the solar system.
Among models of the $r$-process, it has been believed that supernovae are the most
plausible astrophysical site \citep{qi05}.
The explosion is triggered by
the gravitational collapse of massive stars of $M > 10~M_{\odot}~$\citep[e.g.,][]{hashi1995}. Since a proto-neutron star
is formed after the explosion, neutron-rich elements seem to be easily ejected by the supernova
shock. Unfortunately all realistic
numerical simulations concerning the collapse-driven supernovae have
failed to explode the outer layer outside the Fe-core~\citep{jan04}. Therefore,
plausible site/mechanism of the 
$r$-process has not yet been clarified. On the other hand,
explosive nucleosynthesis that produces most elements up to Fe-group nuclei has been calculated under
the assumption that the explosion
is triggered from outside the Fe-core whose location is defined as
the {\it mass cut} \citep{hashi1989,th96};
the calculated abundances from C to Ge are consistent with the supernova observations and the
chemical evolution of galaxies~\citep{tsuji93,tim95}.
However, this situation can not be applied to the $r$-process due to the large electron fraction
$Y_e~(> 0.4)$ and the low entropy distribution above the $mass~cut$. A specific model of neutrino-wind with the very high entropy per baryon ($s/k_{\rm B} \simeq 310$ and $Y_e \simeq 0.45$)
has been suggested to reproduce up to the third peak in the abundance pattern of the
$r$-process~\citep{FRT1999}. It should be noted that
this model includes artificial parameters such as mass loss rates and initial conditions of
hydrodynamical calculations~\citep{sumi2000}.
On the other hand, detailed $r$-process calculations that include
the fission have not been fully performed. If the $r$-process occurs along the paths of the
neutron-drip
line, the fission process should become important~\citep{Panov2001}.

The explosive nucleosynthesis under the jet-like explosion is investigated with use of the 
two-dimensional hydrodynamical code \citep{naga97}, where strong $\alpha$-rich
freezeout region has emerged.
Recently, two-dimensional magnetohydrodynamical (MHD) calculations have been performed under the
various initial parameters concerning the rotation and magnetic field \citep{ko04a,yama04a,ko04b,yama04b,taki04}. The ZEUS-2D code developed by \citet{sn92} has been modified
to include a tabulated equation of state \citep{shen98}, electron captures, and
neutrino transport \citep{ko04a}. Various combinations have been investigated
for the initial ratio of the rotational
energy ($T$) and/or magnetic energy ($E_m$) to the gravitational energy ($W$). In case of
the cylindrical profiles of the rotation and
magnetic field, it is found that the shape of the shock wave
becomes prolate compared to the case without magnetic fields though detailed studies of neutrino
transport must be developed. Furthermore, the confined magnetic
fields behind the shock front push the shock wave strongly.
The aspect ratio of the polar to the equatorial radius in the stalled shock front becomes 1.4
for the initial values of $T/|W| =0.5~\%$ and $E_m/|W| =0.1~\%$.
It is noted that whether the initial models of stellar
evolution include magnetorotational effects~\citep{heg04} or not~\citep{ww95},
no significant differences in the hydrodynamical
features are found.

In the present paper, we will carry out the calculations of the MHD explosion of the 
He-core of 3.3 $M_{\odot}$ whose mass in the main sequence stage is about 13 $M_{\odot}$ star.
Thereafter, we investigate the r-process nucleosynthesis using the
results of MHD calculations. We find new $r$-process site thanks to
a large nuclear reaction network that includes fission; we obtain the region
that produces the $r$-process elements under the low $Y_e$.

In \S 2 the full nuclear reaction network necessary to the $r$-process calculation
is constructed. We describe initial models in \S 3 and give explosion models based on the MHD
simulations.  The results of the $r$-process nucleosynthesis calculations
are presented in \S 4. We summarize and discuss the results in \S 5.

\section{NUCLEAR REACTION NETWORK}

We have developed the nuclear reaction network that had been constructed 
for the rapid-proton capture process \citep{koike04}. 
The network is extended toward the neutron-rich side to the neutron-drip line.
The full network consists of about 4000 nuclear species up to $Z$ = 100. 
We include two body reactions,
i.e., (n,$\gamma$), (p,$\gamma$), ($\alpha$,$\gamma$), (p,n), 
($\alpha$,p), ($\alpha$,n), and their inverses. 
This network contains specific reactions such as 
three body reactions, heavy ion reactions and weak interactions. 
As shown in Table~\ref{tab:ntwk}
we construct two kinds of the
network A and B that consist of different nuclear data set.
For nuclear masses, the experimental data \citep{aw95} is used if available;
otherwise, the theoretical data by mass formula FRDM \citep{moller95} is adopted
in the range $Z\le83$, and/or ETFSI \citep{most} in $8\le Z \le110$.

Most reaction rates are taken from the compilation (REACLIB) of 
\citet{REACLIB1,REACLIB2} that includes experimental and theoretical
       data for the reaction rates and partition functions  
       with use of FRDM (NETWORK A) or ETFSI (NETWORK B). 
Reaction rates for $Z>83$ that are not available in REACLIB are taken from \citet{most}.

The rates on decay channels, $\alpha$-, $\beta^\pm$-decay, and $\beta$-delayed neutron
emission, are taken from JAERI~\citep{JAERI}, 
that includes experimental and theoretical decay rates of nuclei near the stability line.
On the $\beta$-decay rates not available in JAERI,
theoretical rates by \citet{moller} for NETWORK A or those by
\citet{Tachibana} are adopted for NETWORK B.

The same fission data is adopted for both NETWORK A and B.
(a) {\itshape Spontaneous fission}:
Experimental half lives and branching ratios of spontaneous fission are
taken from \citet{JAERI} and \citet{Nudat2.0}. While
theoretical formula of half life \citep[eq. (23)]{KT1975}
with empirical fission barrier \citep{Mamdouh1999,Mamdouh2001}
is adopted for nuclei whose half lives are not known experimentally,
for all nuclei of both $N >$ 155 and $A >$ 240,
the life times of the decay are set to be $10^{-20}$ s \citep{Panov2001}.
(b) {\itshape $\beta$-delayed fission}: 
Branching ratios of $\beta$-delayed fission are taken from \citet{Staudt1992}.
(c) {\itshape Fission yields}: 
Empirical formula \citep[eq. (5)]{KT1975} is adopted about decay products.

Since many charged particles participate in 
the nucleosynthesis during the explosion,
we have included the screening effects for all relevant reactions~\citep{koike04}.
We also use theoretical weak interaction rates that are the function
of the density and temperature \citep{Fuller1,Fuller2}.

\begin{table*}[htbp]
\vspace*{0cm}
\caption{\small $r$-process networks whose range of the mass number
($A$) is determined by the mass formulae of FRDM (upper range in $A$) and ETFSI (lower) for
each element. \label{tab:ntwk}}
\vspace*{0.2cm}
\begin{center}
{\scriptsize
\begin{tabular}{cr|cr|cr|cr|cr}
\hline\noalign{\smallskip}
\multicolumn{1}{c}{Nuclides}&\multicolumn{1}{c}{A}&
\multicolumn{1}{c}{Nuclides }&\multicolumn{1}{c}{A}&
\multicolumn{1}{c}{Nuclides }&\multicolumn{1}{c}{A}&
\multicolumn{1}{c}{Nuclides }&\multicolumn{1}{c}{A}&
\multicolumn{1}{c}{Nuclides }&\multicolumn{1}{c}{A}\\
\noalign{\smallskip}
%\multicolumn{7}{c}{}&\multicolumn{1}{c}{}\\

   \hline
\noalign{\smallskip}
   H &  1 -- 3 &   Sc &39 -- 67& Nb &83 -- 125 & Pm &143 -- 187& Tl &203 -- 263 \\
     &       3 &      &      67&    &      129 &    &       187&    &       255 \\
   He&  3 -- 6 &   Ti &40 -- 70& Mo &86 -- 126 & Sm &144 -- 188& Pb &204 -- 264 \\
     &       6 &      &      72&    &      132 &    &       188&    &       259 \\
   Li& 6 -- 8  &   V  &43 -- 73& Tc &90 -- 129 & Eu &151 -- 189& Bi &209 -- 265 \\
     &      8  &      &      76&    &      133 &    &       193&    &       263 \\
   Be&7 -- 12  &   Cr &44 -- 74& Ru &96 -- 130 & Gd &152 -- 190& Po &210 -- 266 \\
     &      12 &      &      78&    &      136 &    &       196&    &       267 \\
   B & 8 -- 14 &   Mn &46 -- 77& Rh &101 -- 141& Tb &155 -- 198& At &211 -- 269 \\
     &      14 &      &      81&    &       137&    &       197&    &       269 \\
   C &11 -- 18 & Fe & 47 -- 78 & Pd &102 -- 142& Dy &156 -- 212& Rn &215 -- 270 \\
     &      18 &    &       84 &    &       138&    &       202&    &       270 \\
   N &12 -- 21 & Co & 50 -- 81 & Ag &105 -- 149& Ho &161 -- 215& Fr &218 -- 271 \\
     &      21 &    &       85 &    &       147&    &       203&    &       271 \\
   O &14 -- 22 & Ni & 51 -- 82 & Cd &106 -- 150& Er &162 -- 216& Ra &221 -- 272 \\
     &      22 &    &       86 &    &       148&    &       208&    &       272 \\
   F &17 -- 26 & Cu & 56 -- 91 & In &111 -- 155& Tm &167 -- 221& Ac &224 -- 273 \\
     &      26 &    &       89 &    &       149&    &       215&    &       273 \\
   Ne& 7 -- 30 & Zn & 57 -- 94 & Sn &112 -- 156& Yb &168 -- 222& Th &227 -- 274 \\
     &      34 &    &       92 &    &       154&    &       218&    &       274 \\
   Na&20 -- 34 & Ga & 60 -- 95 & Sb &119 -- 162& Lu &173 -- 224& Pa &230 -- 278 \\
     &      37 &    &       97 &    &       161&    &       225&    &       277 \\
   Mg&20 -- 36 & Ge &61 -- 102 & Te &120 -- 164& Hf &174 -- 226& U  &232 -- 280 \\
     &      38 &    &      100 &    &       164&    &       228&    &       280 \\
   Al&22 -- 41 & As &64 -- 103 & I  &123 -- 171& Ta &179 -- 235& Np &235 -- 284 \\
     &      41 &    &      101 &    &       165&    &       229&    &       284 \\
   Si&24 -- 44 & Se & 65 -- 106& Xe &124 -- 180& W  &180 -- 236& Pu &238 -- 287 \\
     &      46 &    &       104&    &       168&    &       232&    &       288 \\
   P &27 -- 45 & Br &68 -- 117 & Cs &129 -- 181& Re &183 -- 239& Am &241 -- 290 \\
     &      49 &    &      117 &    &       181&    &       235&    &       292 \\
   S &28 -- 48 & Kr &69 -- 118 & Ba &130 -- 182& Os &184 -- 240& Cm &244 -- 294 \\
     &      50 &    &      118 &    &       182&    &       236&    &       296 \\
   Cl&31 -- 51 & Rb &74 -- 119 & La &135 -- 183& Ir &189 -- 241& Bk &247 -- 298 \\
     &      51 &    &      119 &    &       183&    &       239&    &       300 \\
   Ar&32 -- 56 & Sr &77 -- 120 & Ce &136 -- 184& Pt &190 -- 242& Cf &250 -- 302 \\
     &      54 &    &      120 &    &       184&    &       243&    &       304 \\
   K &35 -- 55 &  Y &79 -- 121 &Pr  &141 -- 185& Au &195 -- 257& Es &253 -- 306 \\
     &      57 &    &      121 &    &       185&    &       247&    &       308 \\
   Ca&36 -- 62 & Zr &81 -- 122 & Nd &142 -- 186& Hg &196 -- 258& Fm &256 -- 310 \\
      &     60 &    &      124 &    &       186&    &       251&    &       312 \\
  \noalign{\smallskip}
           \hline
\end{tabular}
}
\end{center}
\end{table*}

\section{SUPERNOVA MODELS}

\subsection{Initial Models}

The presupernova model has been calculated from the evolution of He-core of 3.3 $M_\odot$ that
corresponds to 13 $M_\odot$
in the main sequence stage \citep{hashi1995}. The mass of the Fe-core is 1.18 $M_{\odot}$ that
is the smallest Fe-core in massive stars obtained from the stellar evolutionally
calculation with the limitation of the spherical symmetry.
The edge of the Fe-core that has steep density gradient is at $R=8.50\times10^7$ cm from the center.
The mass of the Si-rich layer is
$0.33~M_{\odot}$ and the layer extends to $5.47\times10^8$ cm above the Fe-core.
Since the central density exceeds $10^{10}~\rm g~cm^{-3}$ ($\rho=2.79\times10^{10}~\rm g~cm^{-3}$)
and temperature
$T_9 = 9.04$ in units of $10^9$~K, the Fe-core just begins to collapse.

Initial models for the collapse calculations (precollapse models) are constructed by using the
density and temperature distributions of the original Fe+Si core.
We adopt cylindrical properties of the angular velocity $\Omega$ and the toroidal component of the magnetic field $B_{\phi}$ as follows \citep{ko04a}:

\begin{equation}
	\Omega(X,Z)=\Omega_0\times {X_0^2 \over X^2+X_0^2}\cdot {Z_0^4 \over Z^4+Z_0^4}, \ \ \
	B_\phi(X,Z)=B_0\times {X_0^2 \over X^2+X_0^2}\cdot {Z_0^4 \over Z^4+Z_0^4}
\end{equation}
where $X$ and $Z$ are the distances from the rotational axis and the equatorial plane with
$X_0$ and $Z_0$ being model parameters.
Both $\Omega_0$ and $B_0$ are the initial values at $X=0$ and $Z=0$.
Initial parameters of four precollapse models are given in Table~\ref{tab:ini}. 
The spherically symmetric case is denoted by model 1.
In model 2, the profiles of rotation and magnetic field in the Fe-core 
are taken to be nearly uniform.
We present model 4 as the case having a differentially rapid rotating core and strong magnetic fields. An intermediate example, model 3 between model 2 and 
model 4 is prepared for reference. Since the value of $T/|W|$ is higher compared to that used in
\citet{taki04} by a few percents, we regard the present case of  $T/|W|=0.5$~\%
as rather rapid rotating stars with
the moderate magnetic field.
In all computations, spherical coordinates $(r,\theta)$ are adopted.
The computational region is set to be $0\le r\le4000$~km and $0\le\theta\le\pi/2$, where the
included mass in the
precollapse models amounts to $1.42~M_{\odot}$.
The first quadrant of the meridian section is 
covered with $400 (r)\times30 (\theta)$ mesh points. To get information of mass elements,
five thousand tracer particles are placed within the region of $0.449\le Y_e \le 0.49$
between $0.8~M_{\odot}$ ($r=410$ km) and $1.3~M_{\odot}$ ($r=2200$ km).

\begin{table}[h]
\caption{Initial parameters of precollapse models. \label{tab:ini}}
\begin{center}
%{\scriptsize
\begin{tabular}{lcccccc}
\hline\hline
%\multicolumn{1}{c}{Z}&\multicolumn{1}{c}{A}&
Model & $T/|W|$~(\%) &$E_m/|W|$~(\%) & $X_0^*$ & $Z_0^*$ & $\Omega_0~(\rm s^{-1})$  &$B_0$~(G)\\
\hline
model 1 & 0   & 0    &  0   & 0 & 0   &   0 \\
model 2 & 0.5 & 0.1  &  1   & 1 & 5.2 & 5.4 $\times 10^{12}$\\
model 3 & 0.5 & 0.1  &  0.5 & 1 & 7.9 & 1.0 $\times 10^{13}$\\
model 4 & 0.5 & 0.1  &  0.1 & 1 & 42.9 & 5.2 $\times 10^{13}$\\
\hline\\
\end{tabular}
\tablecomments{$X_0^*=X_0/10^8 \rm~cm$ and $Z_0^*=Z_0/10^8 \rm~cm$.
}
%}
\end{center}
\end{table}

\subsection{Explosion Models}

\begin{figure}[htbp]
\vspace*{0.0cm}
\begin{center}
\includegraphics[width=16cm,keepaspectratio ]{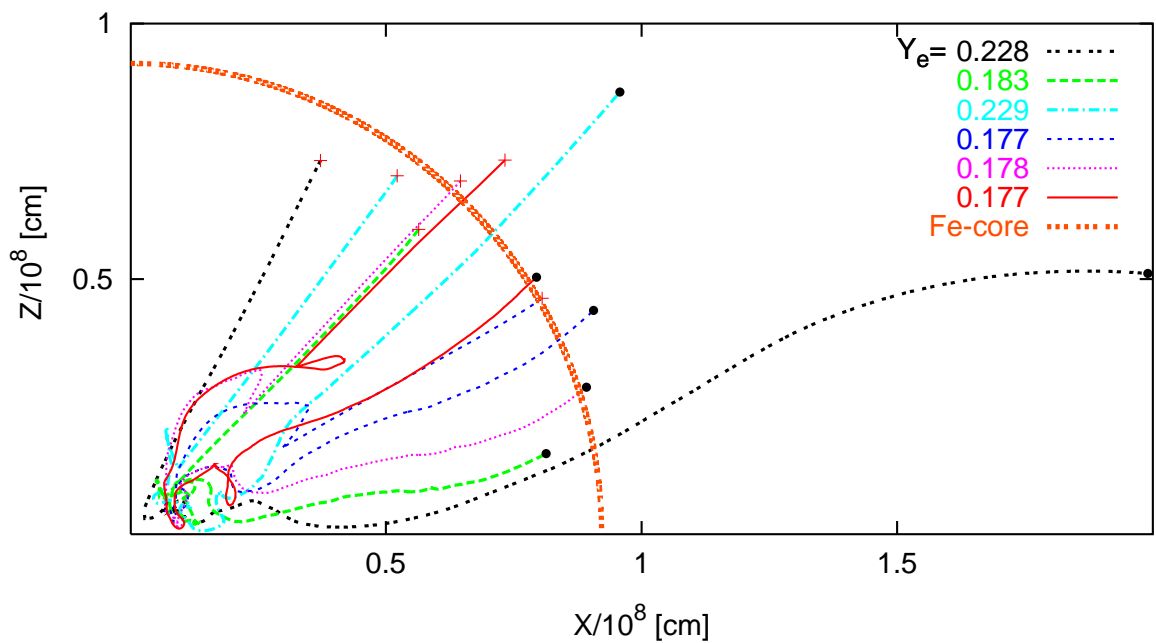}\\
\vspace{0.5cm}
\caption{\small Trajectories of each tracer particle from the initial stage 
(+) to the final stage ($\bullet$) of $565$~ms during the simulation (model~2).
The edge of the Fe-core in the precollapse model is shown by the thick-dotted
line.  The values of $Y_e$ correspond to those in the last stage of NSE for each
tracer particle.
\label{fig:part1}}
\end{center}
\end{figure}

\begin{figure}[htbp]
\begin{center}
\hspace*{1.0cm}
\includegraphics[width=30cm,keepaspectratio ]{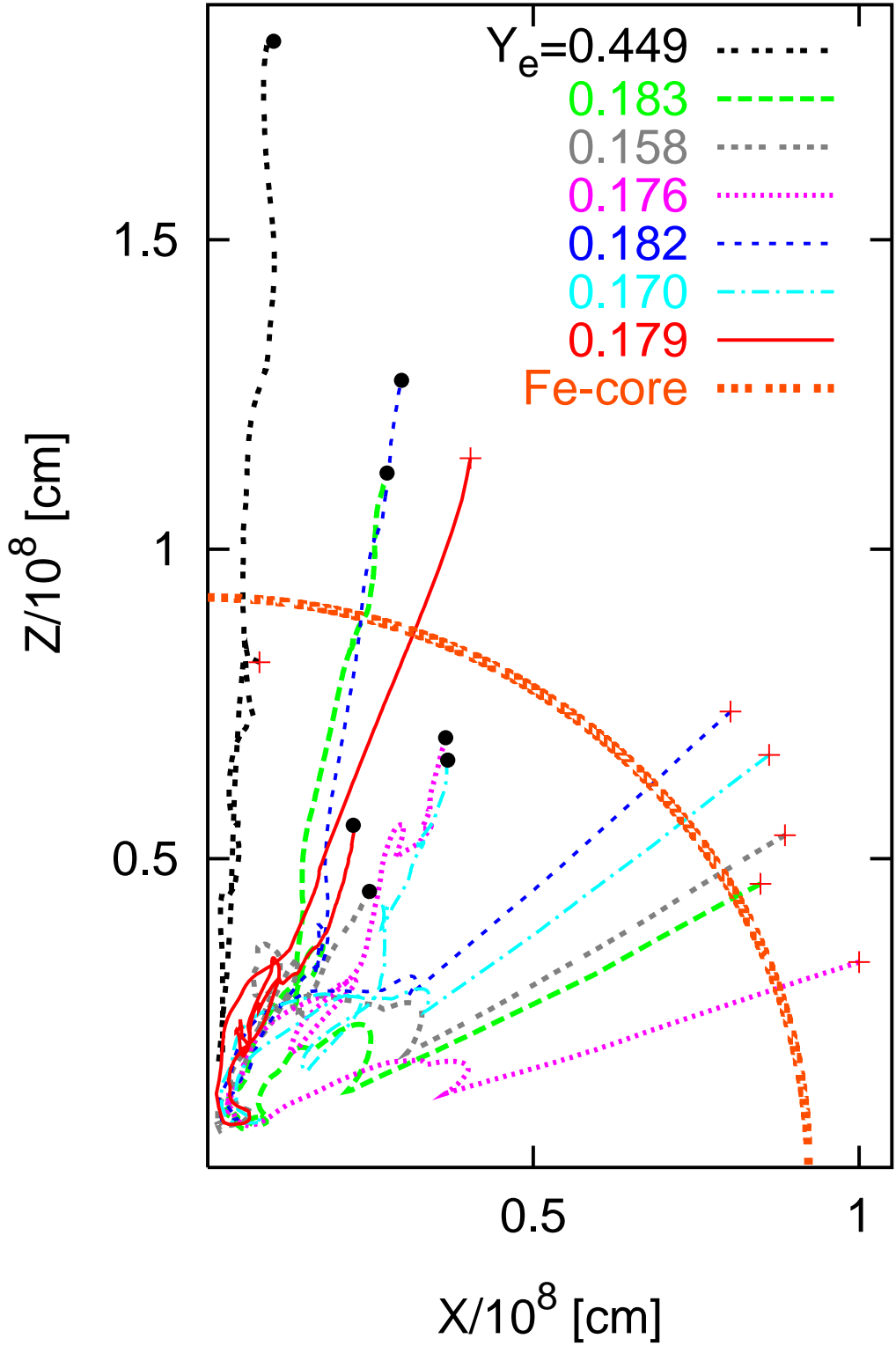}\\
\vspace{0.5cm}
\caption{\small Same as Fig. \ref{fig:part1} but for the final stage ($\bullet$) of $507$~ms
 (model~4).
\label{fig:part2}}
\end{center}
\end{figure}

\begin{figure}
\begin{center}
 \includegraphics[width=9cm,keepaspectratio]{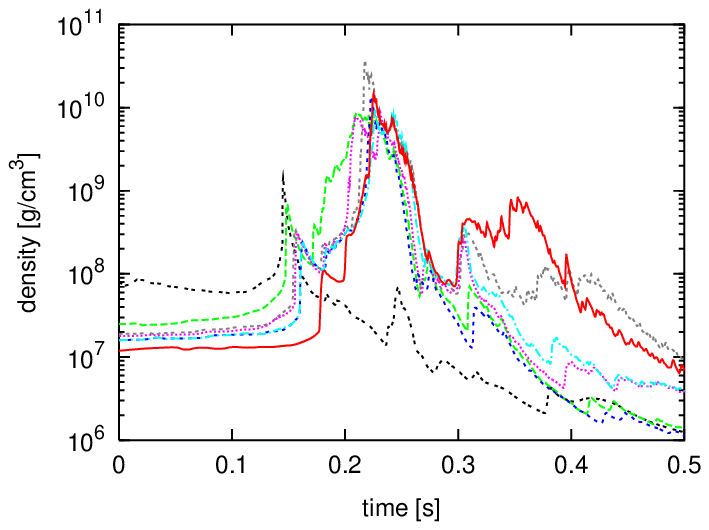}
 \includegraphics[width=9cm,keepaspectratio]{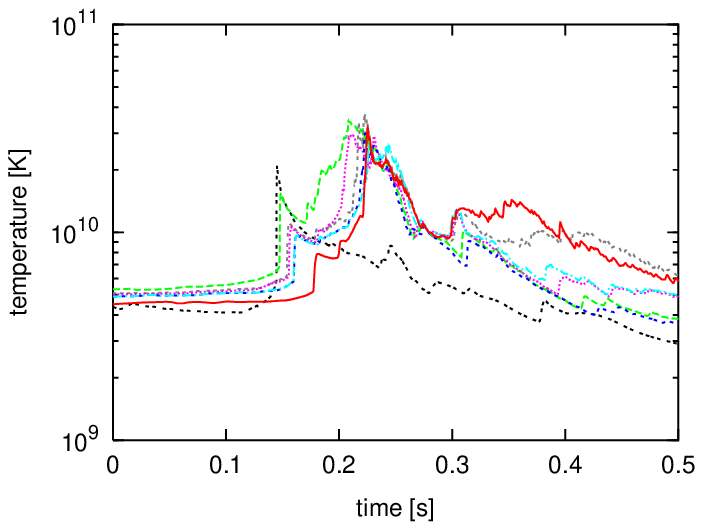}\\
 \includegraphics[width=9cm,keepaspectratio]{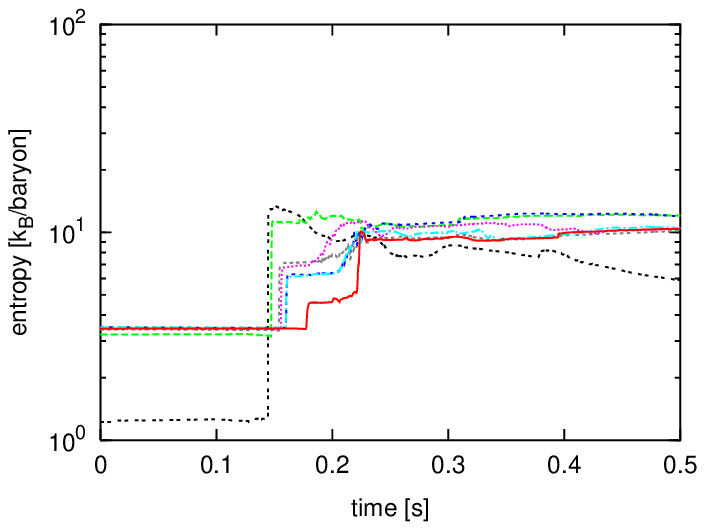}\\
\vspace{0.5cm}
\caption{\small Time evolution of the density, temperature, and entropy
per baryon in $k_{\rm B}$. Each curve corresponds to that in Fig. \ref{fig:part2}
(model~4). \label{fig:rts}}
\end{center}
\end{figure}

We perform the calculations of the collapse, bounce, and the propagation of the shock wave with use of ZEUS-2D in which the
realistic equation of state \citep{shen98} has been implemented by \citet{ko04a}.
We do not include the neutrino transport, since our aim is to clarify the differences in the
nucleosynthesis between spherical and MHD jet explosion.
It is noted that the contribution of the nuclear energy generation
is usually negligible compared to the shock energy.
In Table~\ref{tab:res}, our results of MHD calculations are summarized.
$E_{\rm{exp}}$ is the explosion energy when the shock reaches the edge of the
Fe-core \citep{yama04a}.
In model 3, the explosion is failed due to the specific combination of rotation and
magnetic field between the values of model 2 and model 4; although $E_{\rm{exp}}$ still exists,
the radial distance from the center in the generated shock front at the bounce shrinks gradually
after a few oscillations of the front. Therefore,
it does not always depend on $T/|W|$ and/or $E_m/|W|$ whether the explosion
succeeds or not.
In Figs. \ref{fig:part1} and \ref{fig:part2} trajectories of tracer particles are shown for
some specified values of $Y_e$.
While the jet-like explosion occurs along the equator in model 2 (Fig. \ref{fig:part1}),
collimated jet is emerged
from the rotational axis in model 4 (Fig. \ref{fig:part2}).
Figure \ref{fig:rts} shows the density, temperature, and entropy per baryon in $k_{\rm B}$
of the tracer particles in Fig. \ref{fig:part2}.
We find after the jet-like explosion of model 4 that it remains 1.24 $M_{\odot}$ proto-neutron
star inside the radius
300 km accompanying successive accretion onto the star with 
$dM/dt = 0.43~M_{\odot}~\rm s^{-1}$ at $t=0.35$~s.

\begin{table}[h]
\caption{Calculated quantities that are crucial in the $r$-process. \label{tab:res}}
\begin{center}
\small{
\begin{tabular}{lcccccccc}
\hline\hline
%\multicolumn{1}{c}{Z}&\multicolumn{1}{c}{A}&
Model &$t_b$ & $t_f$& $T/|W|_{f}$ &$E_m/|W|_{f}$ &$ E^*_{\rm{exp}}$& $Y^*_e$
& $M_{\rm{ej}}/M_{\odot}$& $M_{\rm{rej}}/M_{\odot}$ \\
\hline
model 1 & 114 & 424  & 0   & 0     & 2.050   & 0.272 & $1.97\times 10^{-1}$ &$5.64\times 10^{-2}$\\
model 2 & 132 &565& 6.2 & 0.042 & 0.728   & 0.177 &$4.71\times 10^{-2}$  &$9.72\times 10^{-3}$\\
model 3 & 138 & 492  & 6.0 & 0.143  & 0.559   & --    & -- & -- \\
model 4 & 141 &507& 6.0 & 0.130  & 0.306   & 0.158 &$1.65\times 10^{-2}$  &$2.00\times 10^{-3}$\\
\hline\\
\end{tabular}
\tablecomments{$t_b$ indicates the time (ms) at the bounce. The calculations are stopped at the
time $t_f$. The ratios $T/|W|_{f}$ and $E_m/|W|_{f}$ are expressed in \%.
$E^*_{\rm exp}=E_{\rm exp}/10^{51}$~ergs. $Y^*_{\rm e}$ is the value at the last stage of NSE.
$M_{\rm ej}$ is the sum of the ejected tracer particles. $M_{\rm rej}$ is
the ejected mass of the $r$-element for $A\geq 63$.}
}
\end{center}
\end{table}

\section{SIMULATIONS OF THE $R$-PROCESS NUCLEOSYNTHESIS}

\begin{figure}[htb]
\plottwo{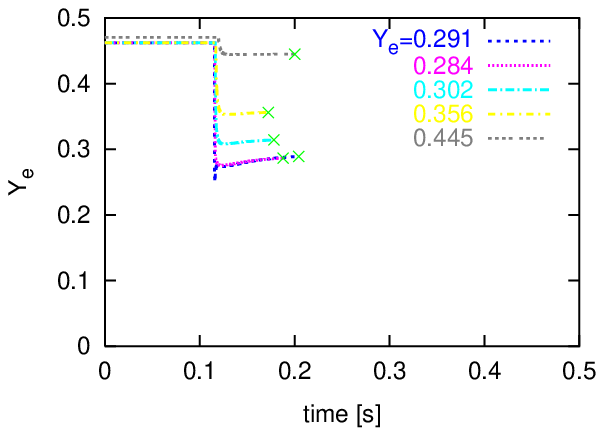}{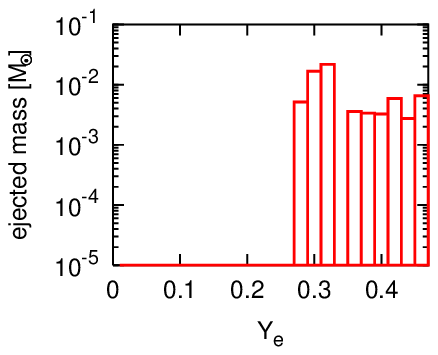}\\
%\vspace{0.5cm}
%\end{figure}
%%%%%%%%%%%%%%%%%%%model~2
%\begin{figure}[htb]
\epsscale{2.2}
\plottwo{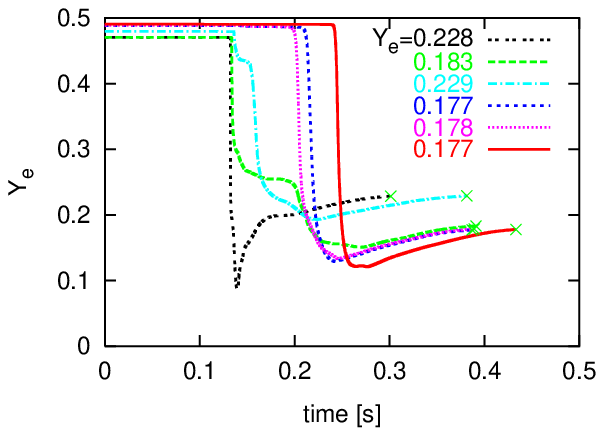}{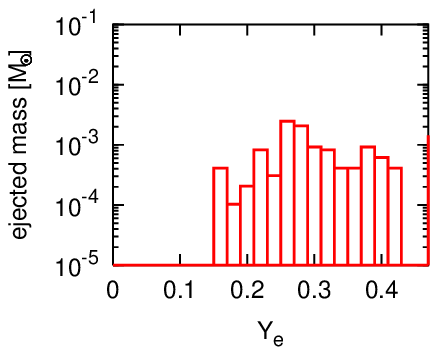}\\
%\vspace{0.5cm}
%Fig.~4b.~---~{Same as Fig.~{\ref{yetm1}}a but for model~2.}
%\label{yetm2}
%\end{figure}
%%%%%%%%%%%%%%%%%%%model~4
%\begin{figure}[htb]
\epsscale{4.8}
\plottwo{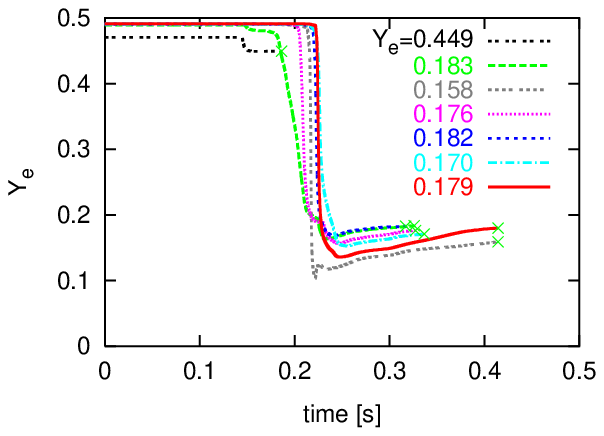}{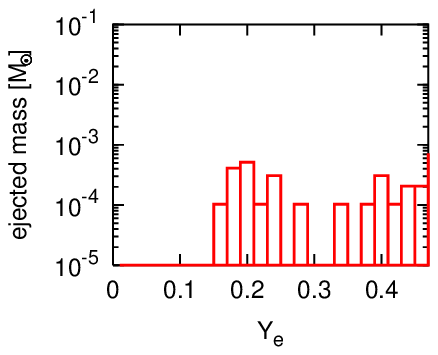}\\
%\vspace{0.5cm}
%Fig.~4c.---~{ Same as Fig.~{\ref{yetm1}}a but for model~4.}
%\label{yetm4}
\caption{\small Time evolutions of $Y_e$~(left panels) and ejected mass vs. $Y_e$~(right
panels) for each tracer particle in model 1 (upper), model 2 (middle), and model 4 (lower).
 \label{yetm}}
\end{figure}

During the explosion,
the temperature exceeds $10^{10}$ K around the layers of the Si+Fe core, where the region of
the nuclear statistical
equilibrium (NSE) is realized as shown in Fig.~\ref{fig:rts}. 
Therefore, we follow the change in $Y_e$ of the ejected tracer particle due to the weak
interactions of electron/positron captures, and $\beta^{\pm}$-decays until the last stage of NSE.
We set this stage to be $T_9 = 9$; afterward the temperature decreases
in time as shown in Fig.~\ref{fig:rts}. The change in $Y_e$ is calculated from the relation

\begin{equation}
{dY_e \over dt}=\sum_{all}\left[\lambda_{+}-\lambda_{-} \right]y_i ,
\end{equation}
where for the abundance $y_i$, $\lambda_+$ consists of the $\beta^{-}$ and positron capture rates, and 
$\lambda_-$ consists of the $\beta^{+}$ and electron capture rates, respectively.
Time evolutions of $Y_e$ relevant to the $r$-process are shown in Fig.~\ref{yetm} (left panels).
The
trajectories of tracer particles with $0.177 \le Y_e \le 0.44$ are depicted in Fig.~\ref{fig:part1}
for model 2 and those with $0.158 \le Y_e \le 0.46$ in Fig.~\ref{fig:part2} for 
model 4, respectively; the values of $Y_e$ indicate those of the last stage of the NSE calculation.
In model 4, the polar region is ejected having rather high value of $Y_e \simeq 0.45$.
The lowest value of $Y_e \simeq 0.16$ is discovered from around the region inclined at 
20 -- 30 degrees from the rotational axis.

Thereafter, using the compositions obtained from the last NSE stage
and the profiles of the density and temperature during the explosion,
we perform the $r$-process nucleosynthesis with the nuclear reaction network described in \S 2.
We remark that after the last stage of the NSE, the changes in $Y_e$ for Fig.~\ref{yetm}
are obtained from
the calculations by the full network. After the time $t_f$, both the temperature and density
are extrapolated to $t=1.2$~s ($T_9 \sim 0.1$) in proportion to ${\rm e}^{-at}$
with $a\sim 1.5$ for
the temperature and $a\sim 4.1$ for the density, respectively \citep{sumi2001}.
We consider that the qualitative 
results of the $r$-process nucleosynthesis do not depend much on the value of $a$.

Figures~\ref{yetm}  (right panels) show the ejected mass in $M_{\odot}$ against $Y_e$
in the range  $0.15 \le Y_e \le 0.46$.
For the spherical explosion, materials with $0.272 \le Y_e \le 0.46$ is ejected.
The ejection for $Y_e < 0.4$ occurs from inside the Fe-core in the range of
$R=660-767$~km. On the other hand, ejection occurs in the direction of the equator
with $0.177 < Y_e < 0.44$ for model 2.
As shown in Fig.~\ref{fig:part2}, materials with $0.158 < Y_e < 0.46$ are emerged for
the jet-like
explosion along the rotational axis (model 4). For both models 2 and 4, the ejection for $Y_e < 0.46$ comes from the Si-rich
layer. We recognize that as against
the spherical explosion, jet-like explosion of model 4 decreases $Y_e$ significantly.

%%%%%%%%%%%%%%%%%%%model~1

\begin{figure}
\begin{center}
 \includegraphics[width=15cm,keepaspectratio]{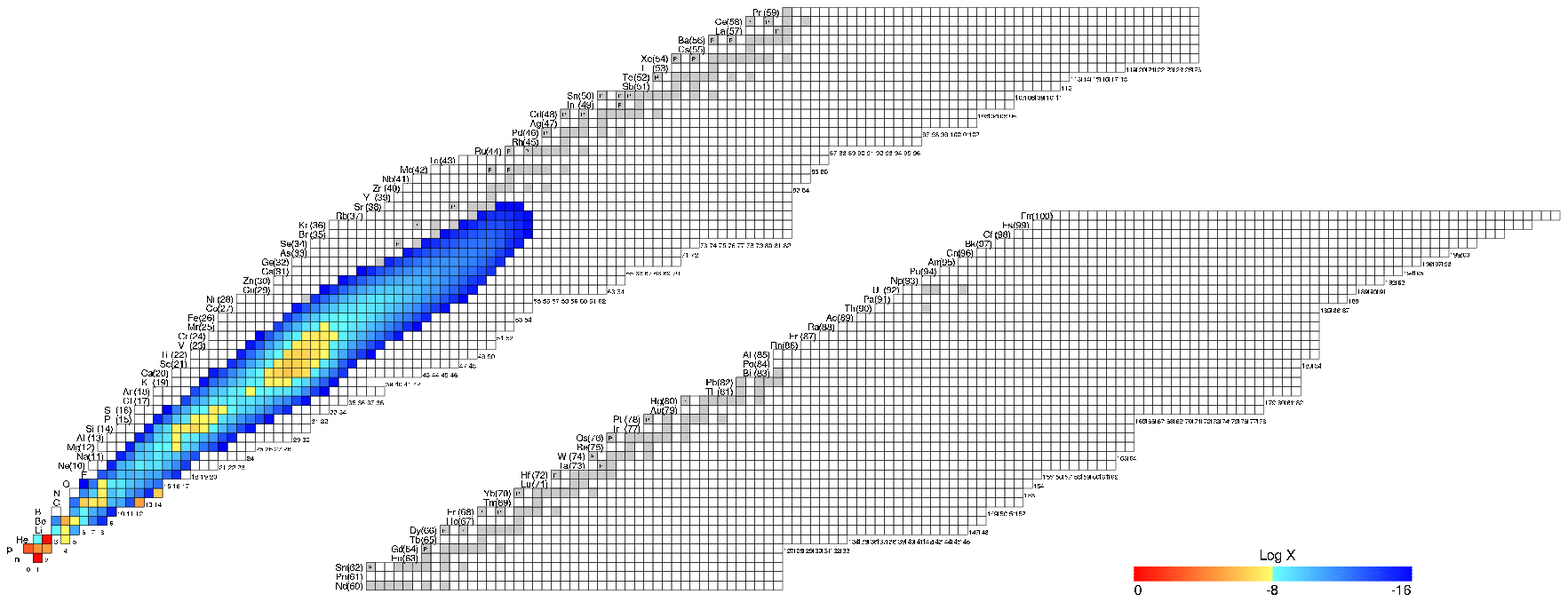}
 \includegraphics[width=15cm,keepaspectratio]{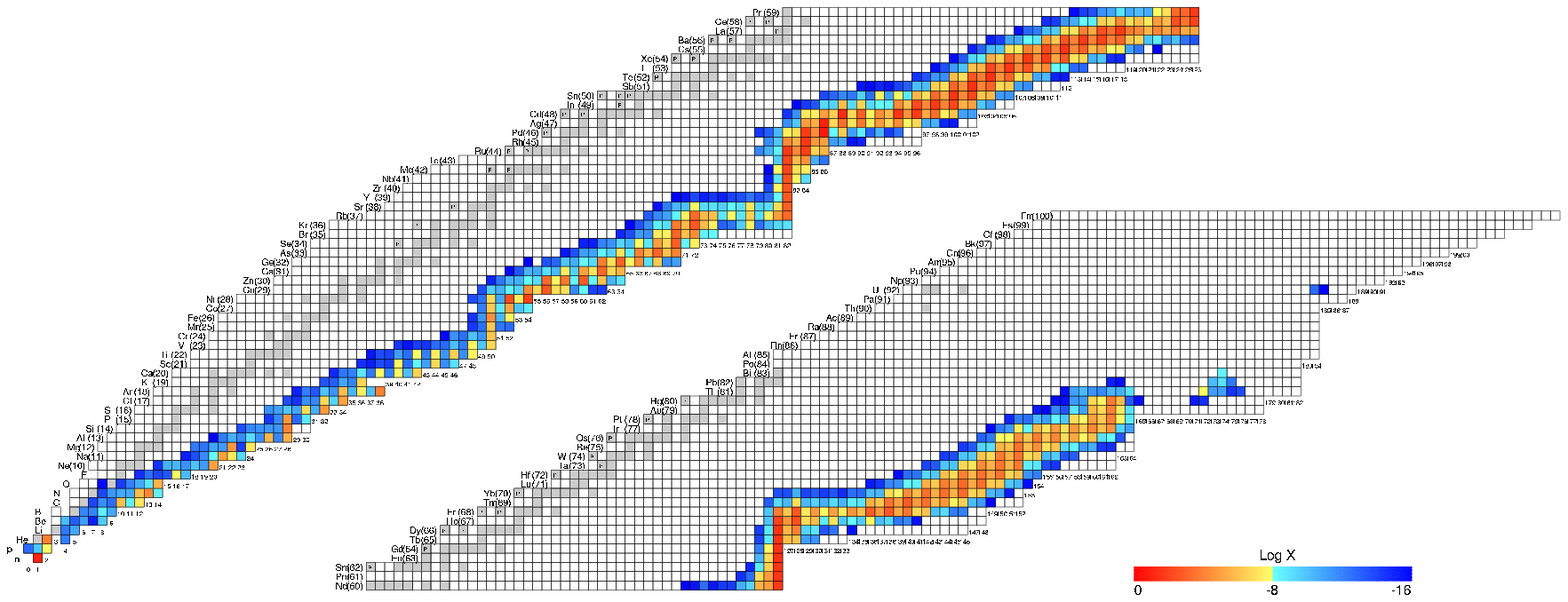}
 \includegraphics[width=15cm,keepaspectratio]{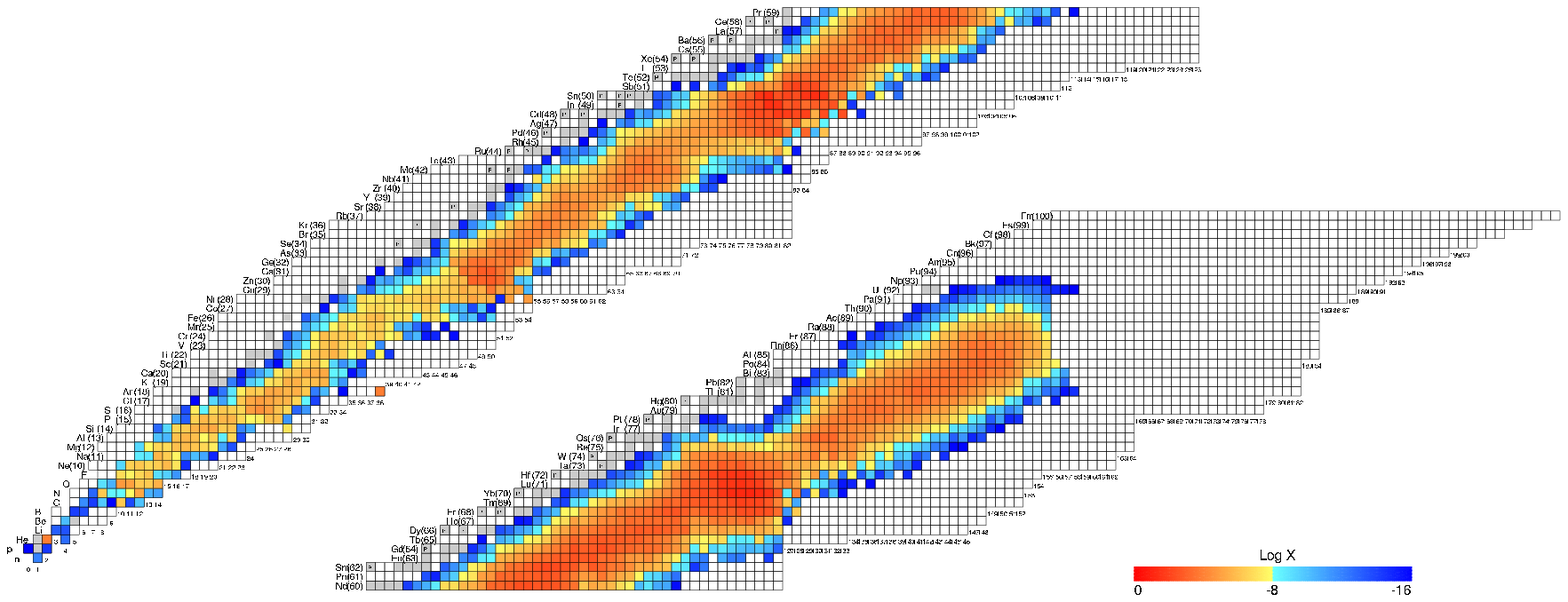}\\
\caption{\small Paths of the $r$-process nucleosynthesis during the jet-like
explosion of model 4; ($t$, $Y_e$) = (0.489~s, 0.161): upper, (1.650~s, 0.312): middle, and (2.760~s, 0.372): lower.
The tracer particle has $Y_e=0.158$ in the last stage of NSE: $T_9 =9$  at $t=0.462$~s from the beginning of the collapse.}
\label{rpro}
\end{center}
\end{figure}

\begin{figure}[htb]
\begin{center}
\hspace{-0.5cm}\includegraphics[width=7.25cm,keepaspectratio]{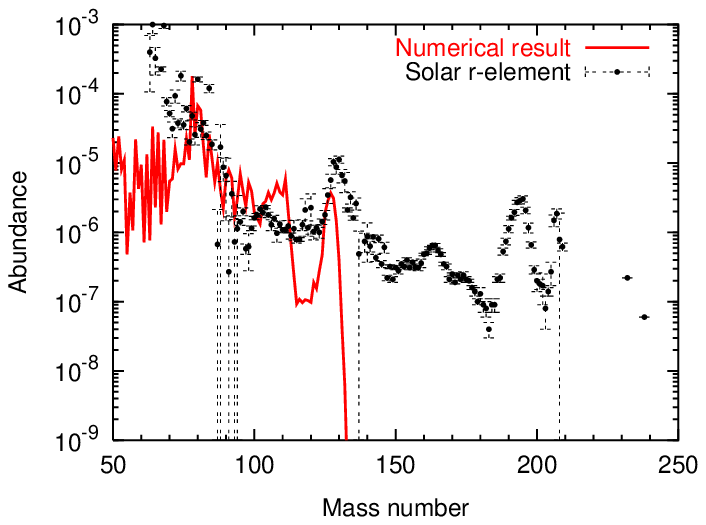}\\
%\vspace*{-0.5cm}\\
%\hspace{0.5cm}Fig.~6a.--- \small{Abundances obtained from the\\
%\hspace{0.5cm}spherical explosion (model 1) with NETWORK A.}
%\label{model1FRDM}
\end{center}
%%\end{figure}
%%
%%\begin{figure}[htb]
\begin{minipage}{.44\linewidth}
\vspace*{-0.5cm}
\includegraphics[width=\linewidth,keepaspectratio]{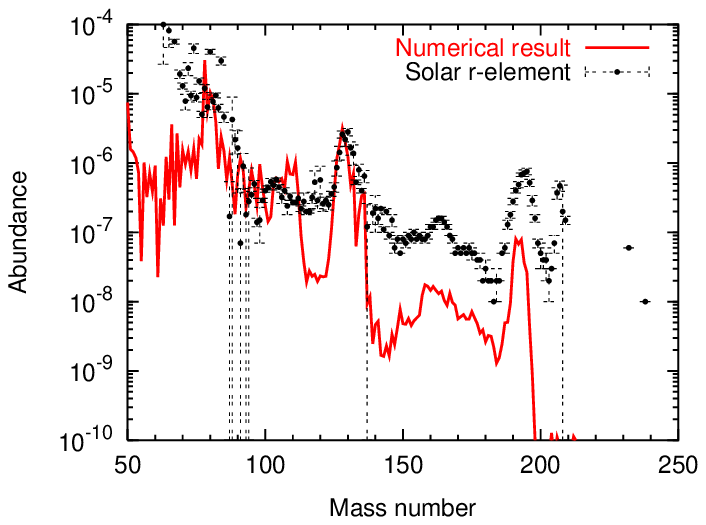}
 \vspace*{-0.5cm}\\
%Fig.~6b.--- \small {Same as Fig.~6a but for model 2.} 
%\label{model2FR}
\end{minipage}
\hspace{1cm}
\begin{minipage}{.44\linewidth}
\vspace{-0.5cm}
\includegraphics[width=\linewidth,keepaspectratio]{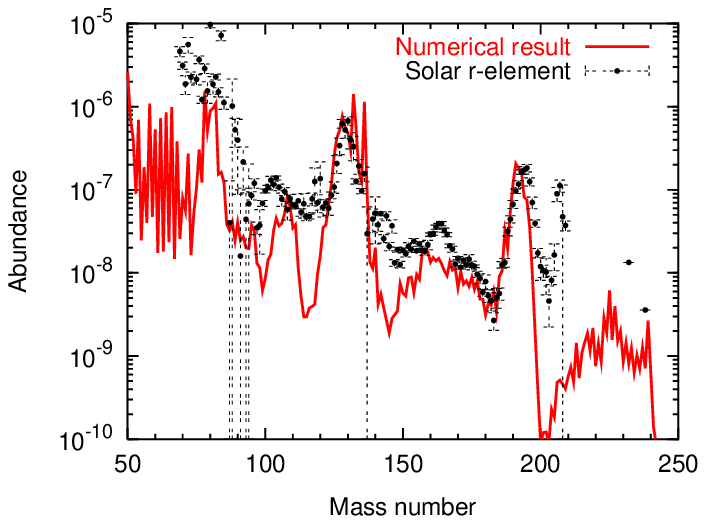}
 \vspace*{-0.5cm}\\
%Fig.~6c.--- \small {Same as Fig.~6a but for model 4.}
%\label{model4FR}
\vspace*{0cm}
 \end{minipage}
\caption{\small Abundances obtained from model 1 (upper), model 2 (lower-left), and
model 4 (lower-right)  with use of NETWORK A.}
\label{modelFRDM}
\end{figure}

\begin{figure}[htb]
\begin{center}
\hspace{-0.5cm}\includegraphics[width=7.25cm,keepaspectratio]{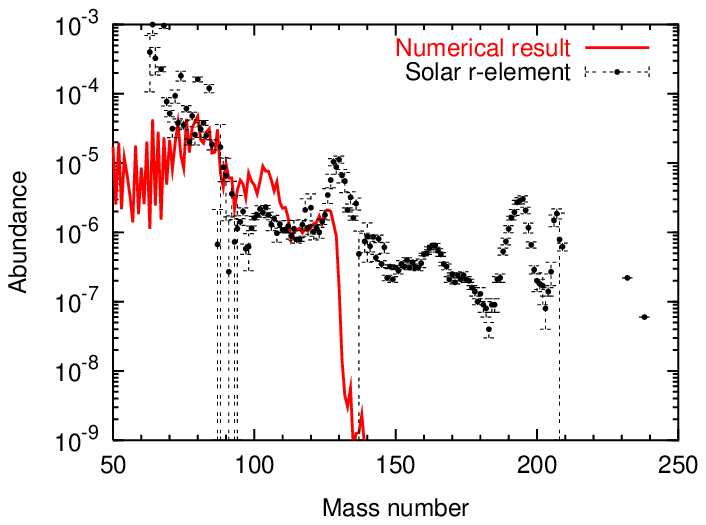}\\
%\vspace*{-0.5cm}\\
%\hspace{0.5cm}Fig.~7a.--- \small {Abundances obtained from the\\
%\hspace{0.5cm}spherical explosion (model 1) with NETWORK B.}
%\label{model1ET}
\end{center}
%%\end{figure}
%%
%%\begin{figure}[htb]
\begin{minipage}{.44\linewidth}
\vspace*{-0.5cm}
\includegraphics[width=\linewidth,keepaspectratio]{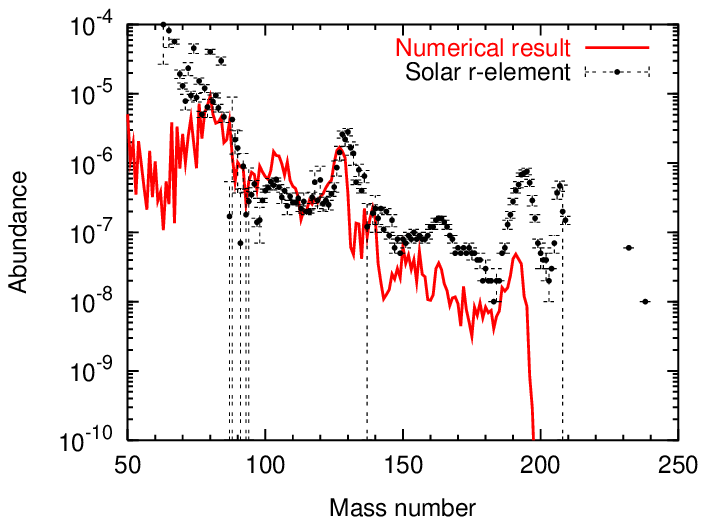}\\
% \vspace*{-0.5cm}\\
%Fig.~7b.--- \small {Same as Fig.~7a but for model 2.} 
%\label{model2ET}
\end{minipage}
\hspace{1cm}
\begin{minipage}{.44\linewidth}
\vspace{-0.5cm}
 \includegraphics[width=\linewidth,keepaspectratio]{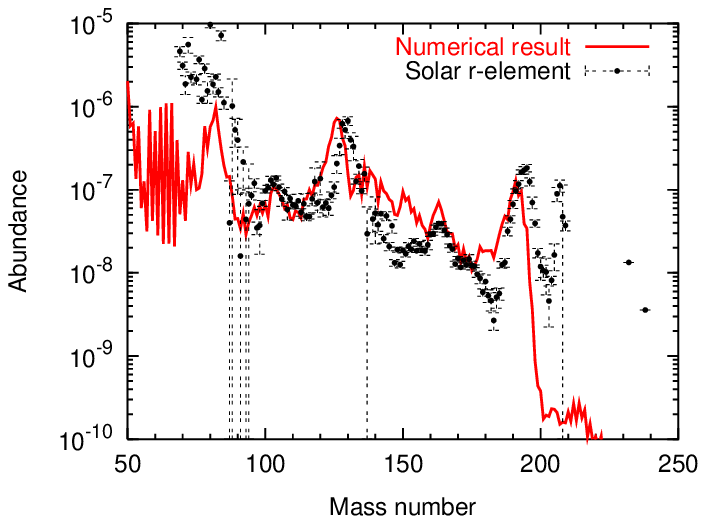}
 \vspace*{-0.5cm}\\
%Fig.~7c.--- \small {Same as Fig.~7a but for model 4.} 
%\label{model4ET}
\vspace*{0cm}
 \end{minipage}
\vspace{0.3cm}\\
%\hspace{3.1cm}Fig.~7.--- Same as Fig.~6 but for NETWORK~B
\caption{\small Abundances obtained from model 1 (upper), model 2 (lower-left), and
model 4 (lower-right)  with use of NETWORK B.}
\label{modelET}
\end{figure}

We calculated the $r$-process nucleosynthesis during the MHD explosion using NETWORKs A and B.
The progress in the $r$-process is shown in Fig.~\ref{rpro} at three epochs in
model 4. 
Since $Y_e$ decreases to 0.158 at the last stage of NSE, the $r$-process paths reach to the
neutron drip line.
Figures~\ref{modelFRDM} and \ref{modelET} show the comparison of the solar $r$-process abundances
with obtained
abundances. Generally, compared to the spherical explosion the jet-like explosion
 results in the increase of the nuclei for $A\ge120$. 
The reproduction of the peaks in the distributions of the $r$-elements depends on
the decrease in $Y_e$ during the early phase of the explosion ($T_9\ge 9$).
In model 1, the produced $r$-elements are not enough to explain even
the second peak in the $r$-process pattern.
In model 2, although the second peak is reproduced well, the amount of the produced
$r$-elements are too small to explain the third peak.
For model 4 we succeed in making the global abundance pattern of the $r$-elements from the
first to the third peak.

Moreover, we find that the fission cycling leads to the normal $r$-process nucleosynthesis
based on (n,$\gamma$) $\rightleftharpoons$ ($\gamma$,n) equilibrium
accompanied with $\beta$-decays for the low $Y_e$ region ($Y_e < 0.2$).
On the other hand, distribution of final products are found to be much sensitive to the mass
formula as seen in the differences between Figs.~\ref{modelFRDM} and \ref{modelET}.
This is because the global $r$-process pattern and the profiles of the peaks in the abundance pattern
depend on the $\beta$-decay rates that have been calculated using the mass formulae
explained in \S 2.

\section{SUMMARY \& DISCUSSION}
 Using ZEUS-2D, we find 
the thermodynamical conditions for the $r$-process to occur. The appropriate condition of the
density, temperature and $Y_e$ can be satisfied under the initial rotation law
and the magnetic field strength of the precollapse model; the jet between the polar and the equatorial radius 
produces the $r$-process elements which emerge from the Si-rich layers above the Fe-core. 
Contrary to the calculation under an artificial explosion energy and a $mass~cut$
located outside the Fe-core in the spherical model~\citep{th96}, we have pursued the calculation
from the collapse to the bounce, and shock wave propagation that attains to the edge of the
Fe-core.
While the spherical explosion reproduces at most to the beginning of the second peak
in the abundance
pattern of the solar $r$-elements, jet-like explosion due to the effects of MHD in the direction of the rotational
axis succeeds in reproducing the third peak.
However, our results depend on the initial parameters for both the strength of the rotation and
magnetic field. For the jet-like explosion from the equator, the reproduction of the $r$-elements
is limited to the second peak. We stress that the $r$-process nucleosynthesis 
can have many variations if the MHD effects play an important role in the supernova explosion.

The shape and the position of peaks depend crucially on the mass formula
and $\beta$-decay rates.
However, it is considered that theoretical data on $\beta$-decay rates has large uncertainty.
Therefore, model 4 can reproduce well the solar $r$-process abundance pattern 
within the uncertainties of $\beta$-decay rates. The $\beta$-decay rates calculated by the gross theory are tend to become
small compared to those obtained by shell model computations. As a consequence, NETWORK A
makes the three peaks more clearly than NETWORK B.
Small abundance peak around $A\sim164$
is not built enough in the present calculations for both networks.
It is hoped that
the information of the $r$-process acquired through
theoretical and observational studies may give severe constraints to the nuclear data.

In the present calculation, the fission process does not play an important
role, since the ejected mass of the very low $Y_e$ region is small. However, production of 
abundances for $A\sim208$ may need the region of $Y_e < 0.15$ as suggested in
Figs.~\ref{modelFRDM} and \ref{modelET}.
Furthermore,
fission should be included in the $r$-process calculation
for the situation of very low $Y_e$ such as a neutron star merger with
$0.05 < Y_e < 0.15$~\citep{FRT1999,ross01}. 

As shown in Table \ref{tab:res} ejected mass of the $r$-elements amounts to
$M_{\rm rej}=2\times10^{-3}~M_{\odot}$ in case of model 4 that is one tenth of
the total ejected mass.
Since the ejected mass of oxygen is $M_{\rm O}~=0.15~M_{\odot}$ for the spherical
explosion of the
$3.3~M_{\odot}$ helium core~\citep{hashi1995}, the ratio $M_{\rm rej}/M_{\rm O} \sim 10^{-2}$
is large by a factor of 100 compared to the corresponding solar ratio of $1.6\times10^{-4}$.
It has been pointed out for the explosion of massive stars the underproduction of
the p-nuclides with respect to oxygen, when normalized to the solar values \citep{rayet}.
However it is found that as far as the case in the 3.3 $M_{\odot}$ helium core, produced p-nuclides
are free from the problem of the underproduction \citep{rayet}. 
Considering uncertainties neglected in the present simulations and the differences 
in the nucleosynthesis between the
spherical and jet-like explosion, the problem of the overproduction
in the $r$-elements should be worth while to examine in detail.

In the present investigation, as the first step we ignore the effects of neutrino transport. 
It is known that neutrinos take off the explosion energy significantly. Therefore, delayed
explosion by neutrino heating is the most promising scope of the supernova explosion. However,
there exists difficulties related to the two dimensional treatment of neutrino
transport \citep{ko05}: even one dimensional simulations that include the detailed
neutrino transport process do not succeed in the explosion~\citep{jan04}. Our purpose has been
to study the effects of the MHD jet on the $r$-process and elucidate the differences 
in the produced $r$-elements from the spherical explosion. 
For the next step, some investigations that include the neutrino transport will be
pursued.

%\section*{Acknowledgements}
%We are grateful to K. Arai for reading the manuscript and giving useful comments.

\end{document}